\documentclass[a4paper, amsfonts, amssymb, amsmath, reprint, showkeys, nofootinbib, twoside]{revtex4-1}
\usepackage[english]{babel}
\usepackage[utf8]{inputenc}
\usepackage[colorinlistoftodos, color=green!40, prependcaption]{todonotes}
\usepackage{amsthm}
\usepackage{mathtools}
\usepackage{physics}
\usepackage{xcolor}
\usepackage{graphicx}
\usepackage[left=23mm,right=13mm,top=35mm,columnsep=15pt]{geometry} 
\usepackage{adjustbox}
\usepackage{placeins}
\usepackage[T1]{fontenc}
\usepackage{lipsum}
\usepackage{csquotes}
\usepackage[pdftex, pdftitle={Article}, pdfauthor={Author}]{hyperref} 
\usepackage{amssymb}
\usepackage{natbib}

\usepackage[LGRgreek]{mathastext} 
\newcommand\tab[1][1cm]{\hspace*{#1}} 

\usepackage{lineno}
\usepackage{setspace}
\usepackage{hyperref}

\begin{document}
\title{Nanostructured LaFeO\textsubscript{3}-MoS\textsubscript{2} for efficient photodegradation and photocatalytic hydrogen evolution}

\author{Subrata Das, Sagar Dutta, Angkita Mistry Tama and M. A. Basith}
    \email[Email address: ]{mabasith@phy.buet.ac.bd}
    \affiliation{Nanotechnology Research Laboratory, Department of Physics, Bangladesh University of Engineering and Technology, Dhaka-1000, Bangladesh.\\ \\ DOI: \href{https://doi.org/10.1016/j.mseb.2021.115295}{10.1016/j.mseb.2021.115295}}


\begin{abstract}
Fabrication of heterogeneous photocatalysts has received increasing research interest due to their potential applications for the degradation of organic pollutants in wastewater and evolution of carbon-free hydrogen fuel via water splitting. Here, we report the photodegradation and photocatalytic hydrogen generation abilities of nanostructured  LaFeO\textsubscript{3}-MoS\textsubscript{2} photocatalyst synthesized by facile hydrothermal technique. Prior to conducting photocatalytic experiments, structural, morphological and optical properties of the nanocomposite were extensively investigated using X-ray diffraction analysis, field emission scanning electron microscopy and UV-visible spectroscopy, respectively. Nanostructured LaFeO\textsubscript{3}-MoS\textsubscript{2} photodegraded $\sim$96\% of rhodamine B dye within only 150 minutes which is considerably higher than that of LaFeO\textsubscript{3} and commercial Degussa P25 titania nanoparticles. The LaFeO\textsubscript{3}-MoS\textsubscript{2} nanocomposite also exhibited significantly enhanced photocatalytic efficiency in the decomposition of a colorless probe pollutant, ciprofloxacin eliminating the possibility of the dye-sensitization effect. Moreover, LaFeO\textsubscript{3}-MoS\textsubscript{2} demonstrated superior photocatalytic activity towards solar hydrogen evolution via water splitting. Considering the band structures and contribution of reactive species, a direct Z-scheme photocatalytic mechanism is proposed to rationalize the superior photocatalytic behavior of LaFeO\textsubscript{3}-MoS\textsubscript{2} nanocomposite.
\end{abstract}


\maketitle

\section{Introduction}


In recent years, semiconductor photocatalysts have manifested promising potential for the production of clean hydrogen (H\textsubscript{2}) fuel via water splitting and for the removal of various toxic pollutants from wastewater under solar illumination \cite{Appavu, Guo, Hu, Yendrapati, Mirzaee, Basith1, Wen}. Notably, perovskite metal-oxide lanthanum ferrite (LaFeO\textsubscript{3}) with relatively narrow band gap ($\sim$2.34 eV) as compared to titania (TiO\textsubscript{2}) photocatalyst, has emerged as a promising photocatalyst because of its visible-light responsive photocatalytic activity, non-toxicity and excellent structural stability under the photochemical reaction conditions \cite{Ismael, Peng, Scafetta}. Moreover, LaFeO\textsubscript{3} has been reported as a promising ferroelectric material possessing spontaneous polarization potential at room temperature \cite{Acharya2010}. Although in recent years, significant amount of research is being conducted for synthesizing nanostructured LaFeO\textsubscript{3} \cite{Mihai, Parida, Tijare}  with high photocatalytic efficiency, it has not been possible yet to employ LaFeO\textsubscript{3} nanoparticles for industrial applications due to their high tendency of agglomeration \cite{Peng}, considerably low charge separation efficiency and consequently, fast recombination rate of photoinduced $e^{-}-h^{+}$ pairs \cite{Ismael, Peng, Mihai}.\\ 
\tab Therefore, currently, numerous research endeavors are being undertaken to alleviate the shortcomings of LaFeO\textsubscript{3} nanoparticles which include surface modification, metals and non-metals doping, engineering heterogeneous nanocomposites and introducing lattice defects \cite{Peng, Ren, Wu2015, Wu2018, Yang}. A number of recent investigations have reported that the photocatalytic efficiency of LaFeO\textsubscript{3} nanoparticles can be substantially increased by forming their Z-scheme heterojunctions with other semiconductor photocatalysts such as g-C\textsubscript{3}N\textsubscript{4}, SnS\textsubscript{2} which have suitable band gap energy and appropriate band edge positions \cite{Wu2018, Acharya2017, Luo}. 
However, according to previous reports, the performance of direct Z-scheme photocatalysts depends largely on the energy band configurations, crystallographic structure, surface states and interfacial features of their constituents which makes it quite challenging to design and develop efficient Z-scheme systems \cite{Zhang}. Therefore, despite showing promising potential, LaFeO\textsubscript{3} based Z-scheme photocatalyst nanocomposites are yet less explored.\\ 
\tab In recent years, molybdenum disulfide (MoS\textsubscript{2}), a layered transition metal dichalcogenide, is being effectively employed in photocatalysis as a promising co-catalyst owing to its unique structure, tunable band gap and intriguing optical properties \cite{Chen, Han, Splendiani}. Especially, mono or few-layer MoS\textsubscript{2} nanosheets have been reported to introduce significant improvement in the photocatalytic performance of various pristine metal oxide nanoparticles e.g. ZnO, YVO\textsubscript{4} etc. by forming type-II or Z-scheme heterojunctions with them \cite{Chen, Qian}. These two-dimensional (2D) ultrathin nanosheets also provide enormous specific surface area and prevent the agglomeration of photocatalyst nanoparticles owing to their favorable dispersion and synergetic effect \cite{Jaramillo, Tama}.  Moreover, MoS\textsubscript{2} nanosheets yield a plenty of exposed active edge sites which can efficiently catalyze the photocatalytic H\textsubscript{2} evolution reaction \cite{Chen, Qin}. Hence, considering the compatible band structures and promising properties, it is highly intriguing to develop a heterojunction between LaFeO\textsubscript{3} nanoparticles and 2D MoS\textsubscript{2} nanosheets \cite{Acharya2020} and extensively investigate its photocatalytic performance.\\
\tab Therefore, in the present investigation, we have synthesized LaFeO\textsubscript{3} nanoparticles via a facile sol-gel method \cite{Hu, Ahsan} and incorporated it with ultrasonically exfoliated few-layer MoS\textsubscript{2} nanosheets by adapting a low-temperature hydrothermal synthesis technique \cite{Tama, Das, Sabarinathan} to investigate comprehensively the photodegradation and photocatalytic hydrogen evolution efficiency of LaFeO\textsubscript{3}-MoS\textsubscript{2} nanocomposite. Interestingly, LaFeO\textsubscript{3}-MoS\textsubscript{2} has outperformed both LaFeO\textsubscript{3} and commercial Degussa P25 TiO\textsubscript{2} nanoparticles in the photodecomposition of a textile dye, rhodamine B (RhB) in aqueous solution as well as solar H\textsubscript{2} generation via water splitting. It is noteworthy that colorful organic dyes e.g. RhB may pose some limitations in evaluating the performance of photocatalysts because of their dye-sensitization effect \cite{Barbero}. 
Hence, we have further investigated the photocatalytic activities of LaFeO\textsubscript{3}-MoS\textsubscript{2} nanocomposite towards the photodegradation of a colorless probe pollutant, ciprofloxacin antibiotic which cannot directly absorb visible light to cause photosensitization \cite{Wen}. Notably, the outcome of this experiment provided additional evidence for the superior photocatalytic performance of as-synthesized LaFeO\textsubscript{3}-MoS\textsubscript{2} nanocomposite towards photodegradation.

\section{Materials and Methods}
\subsection{Materials}
The chemical reagents used were analytical grade lanthanum nitrate hexahydrate [La(NO\textsubscript{3})\textsubscript{3}.6H\textsubscript{2}O], ferric nitrate nonahydrate [Fe(NO\textsubscript{3})\textsubscript{3}.9H\textsubscript{2}O], citric acid [C\textsubscript{6}H\textsubscript{8}O\textsubscript{7}], ammonium hydroxide [NH\textsubscript{4}OH] and molybdenum disulfide [MoS\textsubscript{2}] powder (manufacturer of all reagents is Sigma-Aldrich, Germany).

\subsection{Sample preparation}

\subsubsection{Synthesis of LaFeO\textsubscript{3} nanoparticles}

LaFeO\textsubscript{3} nanoparticles were synthesized by a citrate based sol-gel technique \cite{Hu, Ahsan}. Initially, La(NO\textsubscript{3})\textsubscript{3}.6H\textsubscript{2}O, Fe(NO\textsubscript{3})\textsubscript{3}.9H\textsubscript{2}O and C\textsubscript{6}H\textsubscript{8}O\textsubscript{7} were taken in the molar ratio of 1:1:2 and dissolved in deionized (DI) water under magnetic stirring to form a homogeneous solution. To ensure stability, the pH of this solution was then adjusted to 11 by adding NH\textsubscript{4}OH. Afterward, the resulting mixture was heated at 85 $^{\circ}$C for 3 hours until it turned into viscous gel. The viscous gel was further heated at 95 $^{\circ}$C for 3 more hours to initiate auto combustion. After combustion, the obtained dry gel was ground for 1 hour and finally, calcined for 2 hours at air in a programmable furnace (Carbolite, UK) with a heating and cooling rate of 3 $^{\circ}$C/min to acquire the desired LaFeO\textsubscript{3} nanoparticles by removing volatile substances and carbonaceous materials. We have calcined LaFeO\textsubscript{3} nanoparticles at three different temperatures i.e. 600, 700 and 800 $^{\circ}$C to optimize the calcination temperature in terms of crystallinity and uniform surface morphology. Notably, the crystallite size of LaFeO\textsubscript{3} significantly enhanced when calcined at 900 $^{\circ}$C. Hence, we did not further increase the calcination temperature beyond 800 $^{\circ}$C. For brevity, LaFeO\textsubscript{3} nanoparticles calcined at x $^{\circ}$C will be referred to as LaFeO\textsubscript{3}(x) from now on.

\subsubsection{Synthesis of few-layer MoS\textsubscript{2} nanosheets}

Ultrathin few-layer MoS\textsubscript{2} nanosheets have been prepared directly from their bulk powder with a high yield of 60\% via a facile ultrasound assisted exfoliation technique, which has been described in details elsewhere \cite{Das, Sahoo}. Initially, a 5 mg/mL solution of MoS\textsubscript{2} powder and isopropanol was prepared. The solution was then ultrasonicated for 90 minutes in a simple ultrasonic bath (Powersonic 510, maximum power 100 W, sonication frequency 37 kHz or 80 kHz) by applying a continuous ultrasound wave having constant 100 W power and 80 kHz frequency. After 4 hours of rest, residue bulk powder and sheets with micron-size thickness were removed from the ultrasonicated solution via careful decantation. Finally, the collected supernatant was dried for 5 hours in a universal drying oven (Labtech, Republic of Korea) at 80 $^{\circ}$C to obtain MoS\textsubscript{2} nanosheets.

\subsubsection{Synthesis of LaFeO\textsubscript{3}-MoS\textsubscript{2} nanocomposite}

Low-temperature hydrothermal technique is considered beneficial for synthesizing efficient nanostructured photocatalysts because of its ability of inhibiting grain growth, ensuring small particle size with narrow distribution, enhancing crystallinity and minimizing impurities due to the low processing temperature \cite{Basith1, Shi}. Therefore, in this investigation, we have adopted the facile, low-cost hydrothermal technique to incorporate MoS\textsubscript{2} nanosheets with LaFeO\textsubscript{3}(800) nanoparticles \cite{Chen, Tama}. At first, 1 mmol of LaFeO\textsubscript{3}(800) and 0.05 mmol of MoS\textsubscript{2} were dissolved in 50 mL of DI water (i.e. 5 mole\% MoS\textsubscript{2} to LaFeO\textsubscript{3} content). After stirring rigorously for 2 hours, the mixture was heated at 150 $^{\circ}$C for 14 hours in a Teflon-lined autoclave. After cooling down the obtained suspension to room temperature (RT) naturally, centrifugation and continuous rinsing was carried out using DI water and ethanol. Finally, the mixture was dried at 85 $^{\circ}$C in the oven for 12 hours to get the desired LaFeO\textsubscript{3}-MoS\textsubscript{2} nanocomposite.

\subsection{Characterization techniques}

The powder X-ray diffraction (XRD) spectra of as-prepared nanomaterials were obtained by a diffractometer (PANalytical Empyrean, UK) equipped with Cu-$K_{\alpha}$ radiation (wavelength, $\lambda$ = 1.5418 \AA ). Further, FullProf Suite Software was used for extensive analysis of the experimentally acquired XRD data via Rietveld refinement method \cite{Rodriguez-Carvajal}. Field emission scanning electron microscopy (FESEM) imaging along with the energy dispersive X-ray (EDX) analysis was carried out using a scanning electron microscope (XL30SFEG; Philips, Netherlands) to investigate the surface morphology and elemental composition of the samples. A ferroelectric loop tracer (Marine India, India) coupled with an external amplifier (10 kV) was used to determine the leakage current density as well as polarization vs. electric field (P-E) hysteresis loops of the samples \cite{Basith1}. The driving frequency was maintained at 50 Hz during the measurement of P-E loops. The diffuse reflectance spectra (DRS) of as-prepared materials were recorded at RT by a UV-visible spectrophotometer (UV-2600, Shimadzu, Japan) using barium sulfate powder as the reference (wavelength range 200-1100 nm) \cite{Yendrapati}. A Spectro Fluorophotometer (RF-6000, Shimadzu, Japan) was used to carry out photoluminescence (PL) analysis. 

\subsection{Photocatalytic degradation of dye and antibiotic}

The photocatalytic efficiency of prepared nanomaterials and commercial TiO\textsubscript{2} nanoparticles was evaluated in the photodegradation of RhB dye and ciprofloxacin antibiotic in aqueous medium under solar irradiation \cite{Tama, Wei}. The details of the photocatalytic experiments are provided in electronic supplementary information (ESI). In order to evaluate the precision and experimental validation, all the photocatalytic activity tests were carried out three times at exactly same experimental conditions. 
The experimental values are mean of triplicate measurements and their standard deviations have been presented in the figures as error bars at each data point.

\subsection{Photocatalytic hydrogen evolution}

The photocatalytic H\textsubscript{2} evolution test was performed using a slurry-type photochemical reactor containing 40 mg of photocatalyst homogeneously suspended in 60 mL of DI water \cite{Basith1}. Initially, the system was purged with argon gas for 30 minutes to remove the dissolved oxygen and then irradiated by the 500 W Hg-Xe lamp (Hamamatsu Photonics, Japan).  Under illumination, the evolved H\textsubscript{2} gas was collected at 1 hour interval and then, analyzed and measured in the gas chromatography (GC) device (Shimadzu, Japan) equipped with a thermal conductivity detector (TCD) and a gas analyzer. The GC programming was set up with a reverse polarization so that the hydrogen peaks could be attained in the upward direction. The gas analyzer compared the peak intensities of extracted gases to the peaks observed for standard hydrogen gas. The amount of produced hydrogen gas after 4 hours of irradiation was calculated in $\mu$mol g\textsuperscript{-1} catalyst from this comparison. For comparison, the hydrogen evolution test was also performed using commercially available TiO\textsubscript{2} nanocatalyst under identical experimental conditions. The error analyses were also carried out for these experiments.

\section{Results and discussions}

\subsection{Structural characterization}
\subsubsection{Crystal structure}

\begin{figure}[h!]
 \centering
 \includegraphics[width= 8.5 cm]{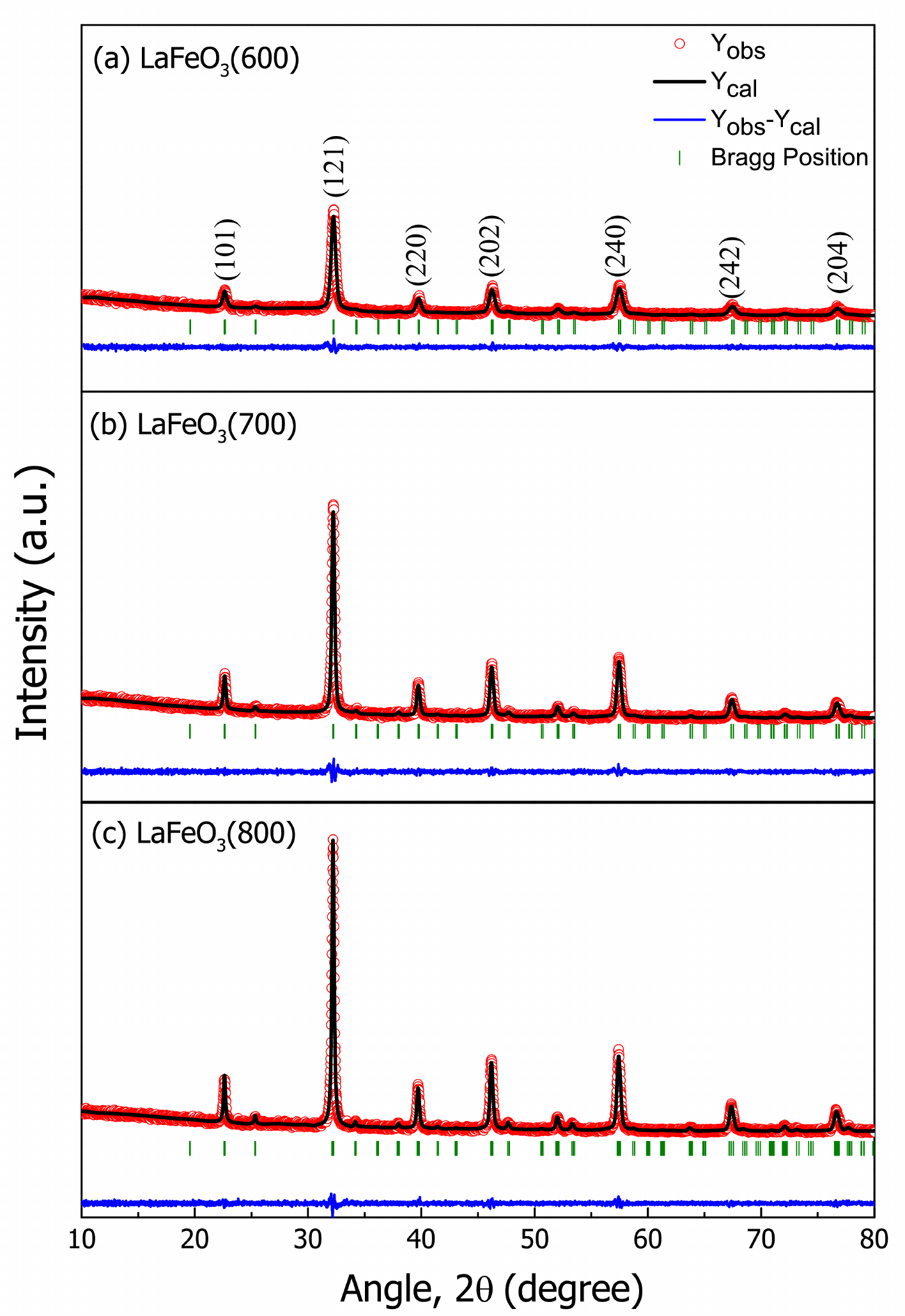}
 \caption{XRD patterns of LaFeO\textsubscript{3} nanoparticles calcined at (a) 600 $^{\circ}$C, (b) 700 $^{\circ}$C and (c) 800 $^{\circ}$C.}
\end{figure}

The crystallographic structure and phase of as-synthesized nanomaterials have been investigated by analyzing their powder XRD spectra via Rietveld refinement method. Fig. 1(a), (b) and (c) display the Rietveld refined powder XRD patterns of LaFeO\textsubscript{3} nanoparticles calcined at 600, 700 and 800 $^{\circ}$C, respectively in the same scale. As shown in the figures, the diffraction peaks exhibited by all these samples can be indexed at (101), (121), (220), (202), (240), (242), and (204) planes (JCPDS card file no 37-1493) which conform to the orthorhombic crystal phase (Pnma space group) of LaFeO\textsubscript{3} perovskite \cite{Acharya2017, Zhao}. No impurity or secondary phases were detected in these XRD spectra, implying that the phase-purity of LaFeO\textsubscript{3} nanoparticles was unperturbed by the varying calcination temperatures. However, the intensity and sharpness of the XRD peaks have considerably enhanced with the increase in calcination temperature which clearly indicates that LaFeO\textsubscript{3}(800) nanoparticles possess higher degree of crystallinity as compared to the other two samples. Our observation agrees well with several previous investigations reporting that higher activation temperature ensures better crystallization for LaFeO\textsubscript{3} nanoparticles \cite{Parida, Wiranwetchayan}. The average crystallite sizes of LaFeO\textsubscript{3} nanoparticles calcined at 600, 700 and 800 $^{\circ}$C are estimated to be about 18, 31 and 45 nm, respectively from the XRD patterns by means of the Scherrer equation \cite{Scherrer}.\\

\tab The XRD spectra of MoS\textsubscript{2} nanosheets and LaFeO\textsubscript{3}-MoS\textsubscript{2} nanocomposite are presented in Fig. 2. As shown in Fig. 2(a), the diffraction peaks demonstrated by MoS\textsubscript{2} nanosheets can be readily indexed to single-phase 2H hexagonal structure (P63/mmc space group) (JCPDS no. 37-1492) \cite{Chen, Liu2014}. The observed small intensity of the XRD peaks of MoS\textsubscript{2} revealed their few-layer sheet-like structure which is qualitatively in good agreement with our previous investigation \cite{Das}. Fig. 2(b) exhibits that all the diffraction peaks of LaFeO\textsubscript{3} nanoparticles along with the most intense peak corresponding to (002) plane of MoS\textsubscript{2} nanosheets are present in the XRD spectrum of LaFeO\textsubscript{3}-MoS\textsubscript{2}, implying the successful formation of the nanocomposite. However, the other diffraction peaks of MoS\textsubscript{2} are not visible in this XRD spectrum because of their relatively low intensity which can be attributed to the small molar ratio of MoS\textsubscript{2} to LaFeO\textsubscript{3} in the nanocomposite. Clearly, the Bragg positions evince the simultaneous existence of two phases corresponding to LaFeO\textsubscript{3} and MoS\textsubscript{2} in the crystal lattice of the nanocomposite. The average crystallite size of LaFeO\textsubscript{3}-MoS\textsubscript{2} nanocomposite has been calculated to be $\sim$36 nm by the Scherrer equation which is smaller than that of pristine LaFeO\textsubscript{3}(800) nanoparticles ($\sim$45 nm).  
\\
\tab Table 1 presents different structural parameters and constituent phases (in wt\%) for as-synthesized nanomaterials. The errors on the last significant digit of the crystallographic parameters are indicated by the numbers in the parentheses. Notably, the calculated lattice parameters of LaFeO\textsubscript{3} and MoS\textsubscript{2} are well consistent with the values reported in literature \cite{Ren, Liu2014}. 
The unit cell volume of LaFeO\textsubscript{3} nanoparticles was observed to nominally enhance with increasing calcination temperature. It is worth noticing that the calculated percentages of the constituent LaFeO\textsubscript{3} and MoS\textsubscript{2} phases, 92.92\% and 7.08\%, respectively,  in the nanocomposite are close to their mole percentages used during synthesis. Moreover, the structural variables corresponding to these two phases remained reasonably unaltered in the LaFeO\textsubscript{3}-MoS\textsubscript{2} nanocomposite, implying that no deformation occurred in the crystallographic structure of individual phases during the preparation of the nanocomposite. For further insight, we have tabulated the calculated atomic coordinates of the samples along with the reliability (R) factors (R\textsubscript{wp}, R\textsubscript{p}, R\textsubscript{Exp} and $\chi^{2}$) of Rietveld refinement in ESI Table S1. The quality of the refinement is evident from the modest R values, in particular the goodness of fit parameter ($\chi^{2}$) of as-synthesized nanomaterials.

\begin{figure}[t!]
 \centering
 \includegraphics[width= 8.5 cm]{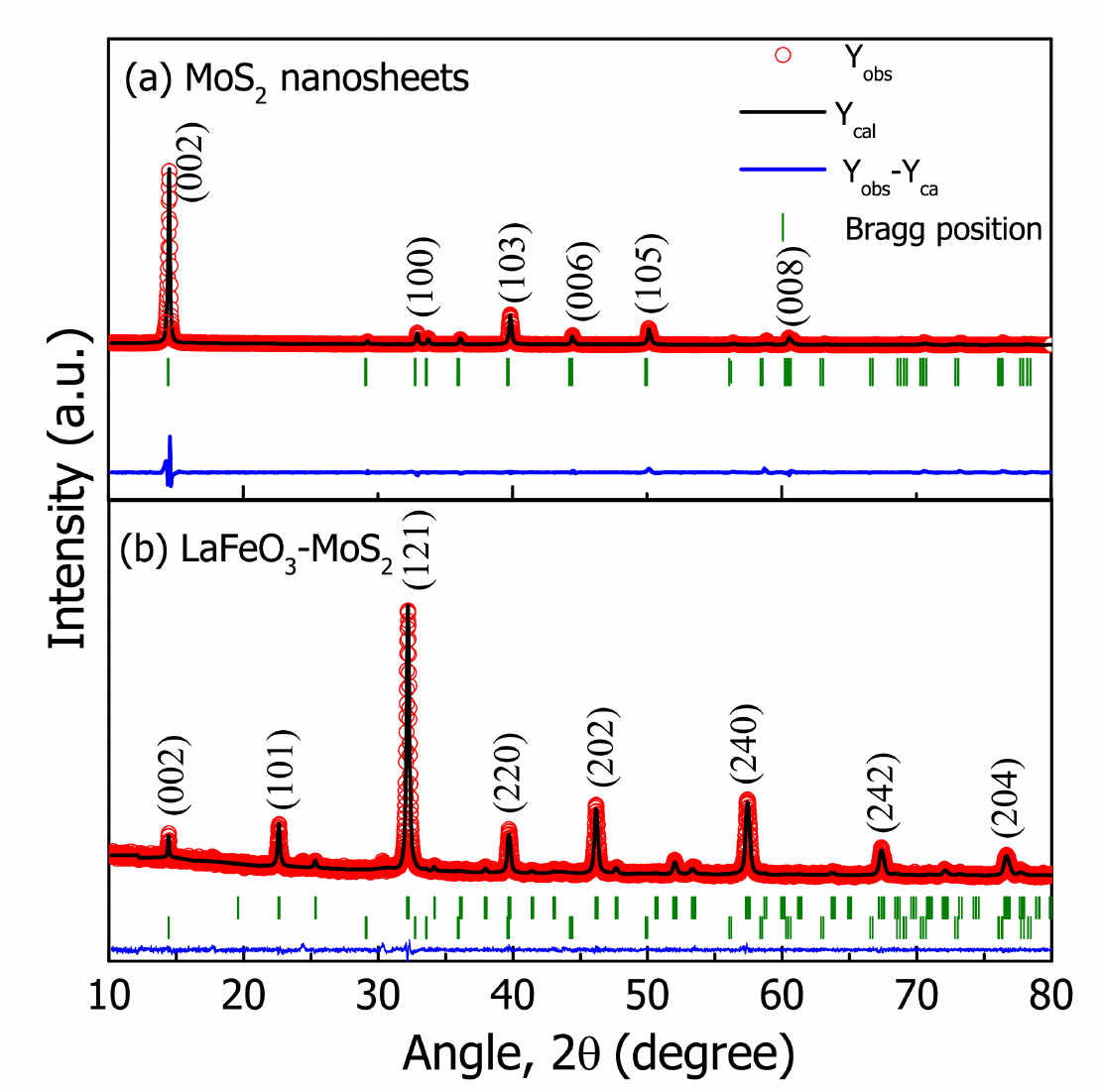}
 \caption{XRD spectra of (a) few-layer MoS\textsubscript{2} nanosheets and (b) LaFeO\textsubscript{3}-MoS\textsubscript{2} nanocomposite. }
\end{figure}

\begin{table*}[t!]
\small
\caption[centering]{Structural parameters of synthesized nanomaterials.}
\begin{tabular*}{\textwidth}{@{\extracolsep{\fill}}llllllll}

\hline
Sample & Constituent & Crystallographic phase (wt\%) & Space group & a (\AA) & b (\AA) & c(\AA) & V (\AA\textsuperscript3)                          \\ \hline
LaFeO\textsubscript{3}(600) &LaFeO\textsubscript{3} & Orthorhombic (100\%) & Pnma &  5.551(0)           & 7.856(0) & 5.551(1)  & 242.10(0)  \\ & & & & \\ 
LaFeO\textsubscript{3}(700) &LaFeO\textsubscript{3} & Orthorhombic (100\%) & Pnma &  5.557(1)           & 7.854(2) & 5.555(1)  & 242.47(0)  \\ & & & & \\ 
LaFeO\textsubscript{3}(800) &LaFeO\textsubscript{3} & Orthorhombic (100\%) & Pnma &  5.569(0)           & 7.853(0) & 5.555(0)  & 242.90(0) \\ & & & & \\ 
 MoS\textsubscript{2} & MoS\textsubscript{2} & Hexagonal (100\%) & P63/mmc & 3.149(0) & 3.149(0) & 12.250(2) & 105.18(0) \\ & & & & \\ 

LaFeO\textsubscript{3}-MoS\textsubscript{2} & LaFeO\textsubscript{3} & Orthorhombic (92.92\%) & Pnma & 5.569(3) & 7.854(6) &  5.554(4) & 242.90(0)  \\
 & MoS\textsubscript{2} & Hexagonal (7.08\%)
& P63/mmc &  3.160(3) & 3.160(3) & 12.277(2) & 106.14(0)
 \\ 
  \\\hline
\end{tabular*}
\end{table*}

\subsubsection{Surface morphology}

\begin{figure}[h!]
 \centering
 \includegraphics[width= 8.5 cm]{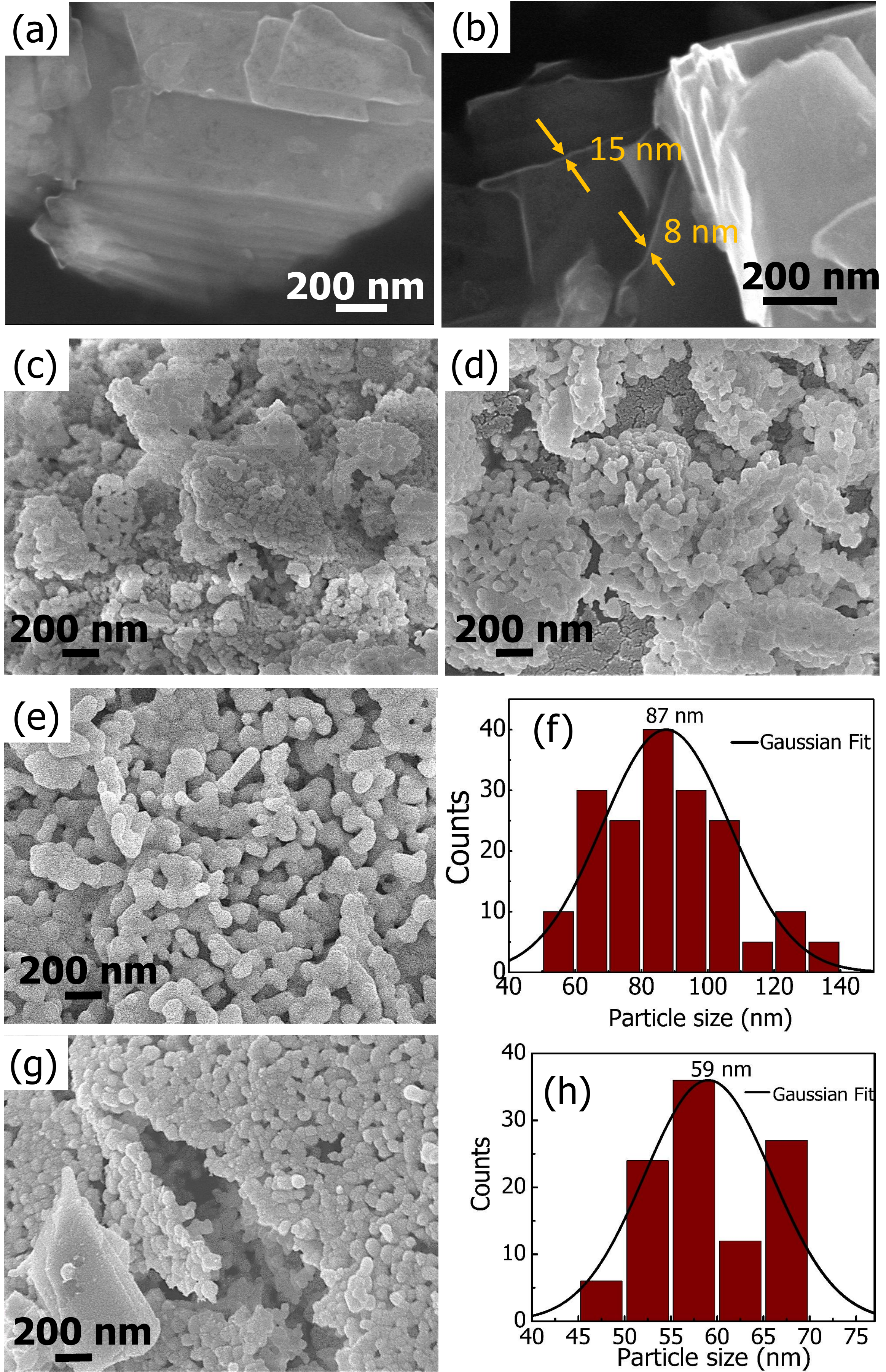}
  \caption{FESEM images of MoS\textsubscript{2} nanosheets : (a) sheet-like structure and (b) nanoscopic thickness. LaFeO\textsubscript{3} nanoparticles calcined at (c) 600 $^{\circ}$C, (d) 700 $^{\circ}$C and (e) 800 $^{\circ}$C. (f) Particle size distribution histogram of LaFeO\textsubscript{3} nanoparticles calcined at 800 $^{\circ}$C. (g) FESEM image and (h) corresponding histogram of LaFeO\textsubscript{3}-MoS\textsubscript{2} nanocomposite.} 
\end{figure}

The surface morphology of as-synthesized samples was investigated by FESEM imaging as shown in Fig. 3. The ultrathin lamellar structure of vertically grown 2D MoS\textsubscript{2} nanosheets is demonstrated in Fig. 3(a). To show the morphological change of MoS\textsubscript{2} due to exfoliation, we have provided two differently magnified FESEM images of bulk MoS\textsubscript{2} powder in ESI Fig. S1. It can be seen that bulk MoS\textsubscript{2} consisted of numerous tightly stacked indistinguishable MoS\textsubscript{2} monolayers which have effectively delaminated to few-layer MoS\textsubscript{2} nanosheets after exfoliation. The thickness of individual sheet can be estimated to be between 8 to 15 nm, suggesting that each nanosheet comprises $\sim$11-21 MoS\textsubscript{2} monolayers (Fig. 3(b)) \cite{Splendiani, Das}. \\
\tab Fig. 3(c), (d) and (e) are the FESEM images of LaFeO\textsubscript{3} nanoparticles calcined at 600, 700 and 800 $^{\circ}$C, respectively. As evident from the figures, at lower calcination temperatures (600-700 $^{\circ}$C), the particles are irregular in shape with non-homogeneous and porous surface morphology. However, with a further increment in calcination temperature to 800 $^{\circ}$C, the surface morphology was considerably improved and the degree of agglomeration was reduced to a moderate extent albeit at the cost of slightly increased particle size. Such reduction in agglomeration might be due to the enhancement of total specific surface area and the free energy of the system, whereas the increment in particle size can be imputed to the increased crystal growth rate of the particles at elevated temperature \cite{Ahsan, Wiranwetchayan, Pang}. The size distribution histogram in Fig. 3(f) demonstrates that the particle size of nanostructured LaFeO\textsubscript{3}(800) lies in the range of $\sim$50-140 nm with an average of $\sim$87 nm.\\ 
\tab Fig. 3(g) and (h) show the FESEM image and corresponding size distribution histogram of  LaFeO\textsubscript{3}-MoS\textsubscript{2} nanocomposite. As can be clearly seen, LaFeO\textsubscript{3} nanoparticles are successfully assembled on the surface of the 2D MoS\textsubscript{2} nanosheets with almost no agglomeration. The satisfactorily homogeneous and non-porous surface of the as-prepared nanocomposite reveals that the growth of LaFeO\textsubscript{3} was much more controlled in the presence of MoS\textsubscript{2}. The improved growth of nanoparticles can be attributed to the presence of LaFeO\textsubscript{3}-anchoring hydroxyl group over the MoS\textsubscript{2} surface \cite{Das, Vattikuti}. Additionally, we anticipate that the 2D MoS\textsubscript{2} nanosheets have provided sufficient surface sites for anchoring LaFeO\textsubscript{3} to prevent aggregation \cite{Han, Tama}. Noticeably, the shape of the particles has become nearly spherical and reasonably uniform after incorporating MoS\textsubscript{2}. Moreover, the particle size distribution of the nanocomposite was found to be considerably narrower ($\sim$45-70 nm) with smaller average size ($\sim$59 nm) as compared to LaFeO\textsubscript{3} nanoparticles which is qualitatively consistent with the outcome of our XRD analysis. However, for both the nanomaterials, the SEM images exhibit slightly larger average particle size as compared to their crystallite size calculated by Scherrer equation which is a common phenomenon according to other reported works \cite{Liu2009}. Finally, based on the observed surface morphology, the as-synthesized LaFeO\textsubscript{3}-MoS\textsubscript{2} nanocomposite can be expected to exhibit better adsorption and photocatalytic efficiency than LaFeO\textsubscript{3} nanoparticles owing to their higher degree of uniformity in shape and size as well as increased surface to volume ratio \cite{Basith1, Peng, Ren}.

\subsubsection{Elemental composition}

The Table S2 inserted in ESI presents the elemental composition of LaFeO\textsubscript{3}(800) nanoparticles obtained both via EDX analysis and theoretical calculation \cite{Wiranwetchayan}. As can be seen in Table S2 of ESI, the experimentally obtained mass and atomic percentages of desired elements La, Fe and O are in excellent agreement with the theoretical values, providing further evidence for the successful synthesis of LaFeO\textsubscript{3} nanoparticles without impurities. Moreover, the La:Fe atomic ratio in the as-prepared sample has been found to be 1:1 which is same as the molar ratio of precursors La(NO\textsubscript{3})\textsubscript{3} and Fe(NO\textsubscript{3})\textsubscript{3} used in the reaction solution. This indicates that the entire metal contents of the precursors were successfully carried over to the final sample under the synthesis conditions. It is worth mentioning that no carbon content was detected in these nanoparticles which can be ascribed to the complete decomposition of carbonaceous compounds from the gel by the calcination process at optimized temperature i.e. 800 $^{\circ}$C. ESI Fig. S2 demonstrates the EDX spectra of LaFeO\textsubscript{3}(800) nanoparticles and LaFeO\textsubscript{3}-MoS\textsubscript{2} nanocomposite. As shown in the Fig. S2 of ESI, along with the peaks of La, Fe and O atoms, the EDX spectrum of LaFeO\textsubscript{3}-MoS\textsubscript{2} contains additional peaks of Mo and S at around 2 keV, confirming the successful formation of the nanocomposite.

\subsection{Electric characterization}


We have determined the ferroelectric polarization (P) as well as the leakage current density (J) of the as-prepared nanomaterials for varying electric fields (E) of up to $\pm$ 60 kV/cm. The driving frequency was maintained at 50 Hz while increasing the applied electric field.  As demonstrated in ESI Fig. S3, the leakage current density of the LaFeO\textsubscript{3}-MoS\textsubscript{2} nanocomposite was considerably smaller as compared to that of the LaFeO\textsubscript{3}(800) nanoparticles. The high leakage current of the nanoparticles can be imputed to their surface porosity as was evident from SEM analysis \cite{Van}. We anticipate that the leakage current has been reduced for the nanocomposite due to their non-porous homogeneous surface morphology. ESI Fig. S4(a) and (b) demonstrate the P-E hysteresis loops of LaFeO\textsubscript{3}(800) nanoparticles and LaFeO\textsubscript{3}-MoS\textsubscript{2} nanocomposite, respectively. As can be observed, the P-E loop of LaFeO\textsubscript{3} is more dominated by leakage current in comparison with the nanocomposite \cite{Deng}. Freely movable charges might be considered responsible for their cigar-shaped unsaturated hysteresis loop \cite{Basith1, Deng}. Interestingly, after incorporating MoS\textsubscript{2} nanosheets, the P-E loops tended towards saturation at high electric fields, indicating the improved ferroelectric behavior of the nanocomposite \cite{Deng, Scott}. The remanent polarization and coercive field of the nanocomposite enhanced with increasing applied voltage since stronger electric fields contribute more to the reversal of ferroelectric domains due to higher driving power \cite{Wu2012}. Moreover, it is noteworthy that the breakdown voltage of LaFeO\textsubscript{3}-MoS\textsubscript{2} (50 kV) was significantly higher than that of LaFeO\textsubscript{3}(800) nanoparticles (10 kV). The interaction between the free charges of 2D MoS\textsubscript{2} and bound charges of LaFeO\textsubscript{3} can be associated with the superior ferroelectric behavior of the nanocomposite \cite{Jin} which is an important characteristic for improving photochemical reactivity \cite{Khan}. The improved ferroelectric behavior of LaFeO\textsubscript{3}-MoS\textsubscript{2} nanocomposite can be anticipated to facilitate spatially selective adsorption of dye and $H_{2}O$ molecules on their surface by creating electric field in the surrounding medium due to surface polarization \cite{Basith1, Acharya2010, Cui}. To be specific, different regions of the surface may experience different extents of band bending depending on the surface polarization and thus, promote different carriers to move to spatially different locations. This phenomenon might result in spatially selective adsorption at surface and facilitate surface redox reactions. Additionally, the internal space charge layer in this ferroelectric nanocomposite may considerably increase the charge separation and migration efficiency by driving the $e^{-}-h^{+}$ pairs in opposite direction and thus, suppress their recombination \cite{Acharya2010, Cui} and improve photocatalytic performance.\\ 

\subsection{Optical characterization}
\subsubsection{Optical band gap}

\begin{figure}[h!]
 \centering
 \includegraphics[width= 0.5\textwidth]{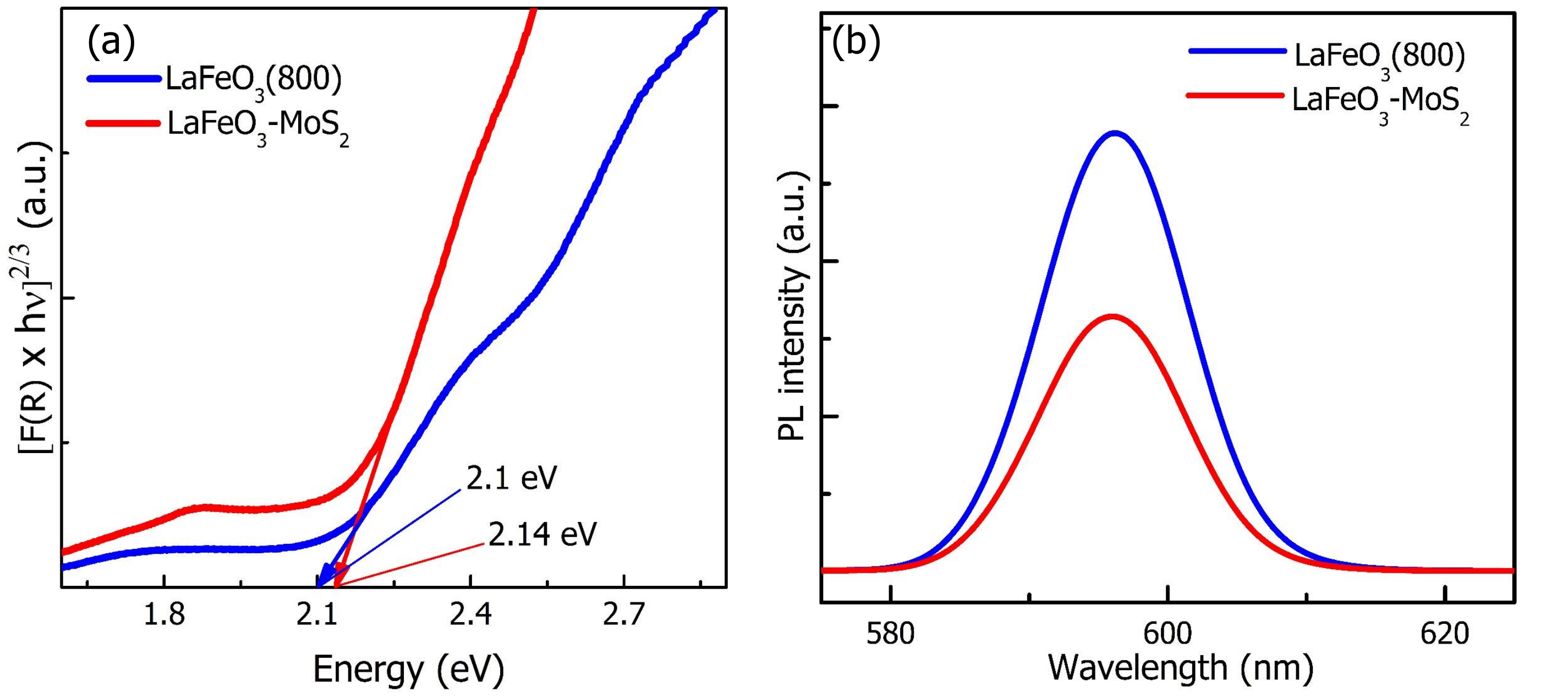}
 \caption{(a) Tauc plots for band gap estimation and (b) steady-state photoluminescence (PL) spectra (excitation wavelength= 300 nm) of LaFeO\textsubscript{3} nanoparticles calcined at 800 $^{\circ}$C and LaFeO\textsubscript{3}-MoS\textsubscript{2} nanocomposite.}
\end{figure}

Optical band gap energy of semiconductors is considered as a crucial factor to assess their photocatalytic activity. Therefore, we have measured the UV-visible DRS of as-prepared photocatalysts and estimated their band gaps using Tauc model \cite{Scafetta} based on eqn. (1)- 
\begin{equation}
    F(R) \times h\nu=A(h\nu-E_{g})^{n}
\end{equation}

where F(R) is a parameter calculated from DRS by Kubelka–Munk function which is proportional to the absorption co-efficient \cite{Basith1} and $h\nu$, A, $E_{g}$ and n signify the energy of photons, proportionality, optical band gap and the electronic transition type, respectively. Notably, Scafetta \textit{et al.} have reported direct-forbidden transition in bulk LaFeO\textsubscript{3} and hence, we have considered n=3/2 to obtain the desired Tauc plots \cite{Scafetta}. Moreover, from the steady-state photoluminescence (PL) spectroscopy, we observed evidence in support of the direct band gap of the synthesized nanomaterials which will be discussed later. As shown in Fig. 4(a), the direct optical band gaps of as-prepared LaFeO\textsubscript{3}(800) and LaFeO\textsubscript{3}-MoS\textsubscript{2} have been estimated to be 2.1 and 2.14 eV, respectively from the abscissa intercepts of the tangents to the linear region of the plots \cite{Acharya2017}. The nominal difference between the observed band gap values can be associated with the synergetic interaction developed due to incorporation of MoS\textsubscript{2} nanosheets \cite{Tama}. However, for both as-prepared nanomaterials, the band gaps are smaller than that of bulk LaFeO\textsubscript{3} ($\sim$2.34 eV) \cite{Scafetta}, suggesting their superior potential to exploit broader range of the solar spectrum and exhibit enhanced visible light-responsive photocatalytic activities.

\subsubsection{Photoluminescence (PL) spectra at steady-state}

The charge-carrier recombination phenomenon in semiconductor photocatalysts can be predicted to a certain extent by performing steady-state photoluminescence (PL) analysis. Generally, an intense PL peak indicates high radiative recombination rate of photogenerated $e^{-}-h^{+}$ pairs. Fig. 4(b) presents the steady-state PL spectra of as-prepared LaFeO\textsubscript{3}(800) and LaFeO\textsubscript{3}-MoS\textsubscript{2} for an excitation wavelength of 300 nm \cite{Wu2018}. From the position of the PL peaks, the band gaps of LaFeO\textsubscript{3} and LaFeO\textsubscript{3}-MoS\textsubscript{2} can be estimated as 2.06 and 2.08 eV, respectively. These values closely match with the band gaps obtained by UV-visible spectroscopy (Fig. 4(a)), providing evidence for the direct transition in these nanomaterials. Moreover, it can be clearly observed that the PL peak intensity of LaFeO\textsubscript{3}-MoS\textsubscript{2} nanocomposite is lower than that of LaFeO\textsubscript{3} nanoparticles which suggests that comparatively smaller number of photogenerated electrons and holes have radiatively recombined in the nanocomposite. Such observation implies that incorporation of few-layer MoS\textsubscript{2} nanosheets can effectively suppress the radiative recombination of charge-carriers. Notably, our observation agrees well with a number of previous studies reporting that MoS\textsubscript{2} nanosheets can substantially quench PL of different metal-oxide photocatalysts by developing heterostructured nanocomposite \cite{Tama, Qin}. However, it should be mentioned that quenched PL peak does not always correspond to enhanced photocatalytic activity since information about the non-radiative recombination cannot be extracted from PL characterization.

\subsection{Photocatalytic degradation of rhodamine B (RhB) dye}

\begin{figure*}[t!]
 \centering
 \includegraphics[width=16 cm, height=19 cm]{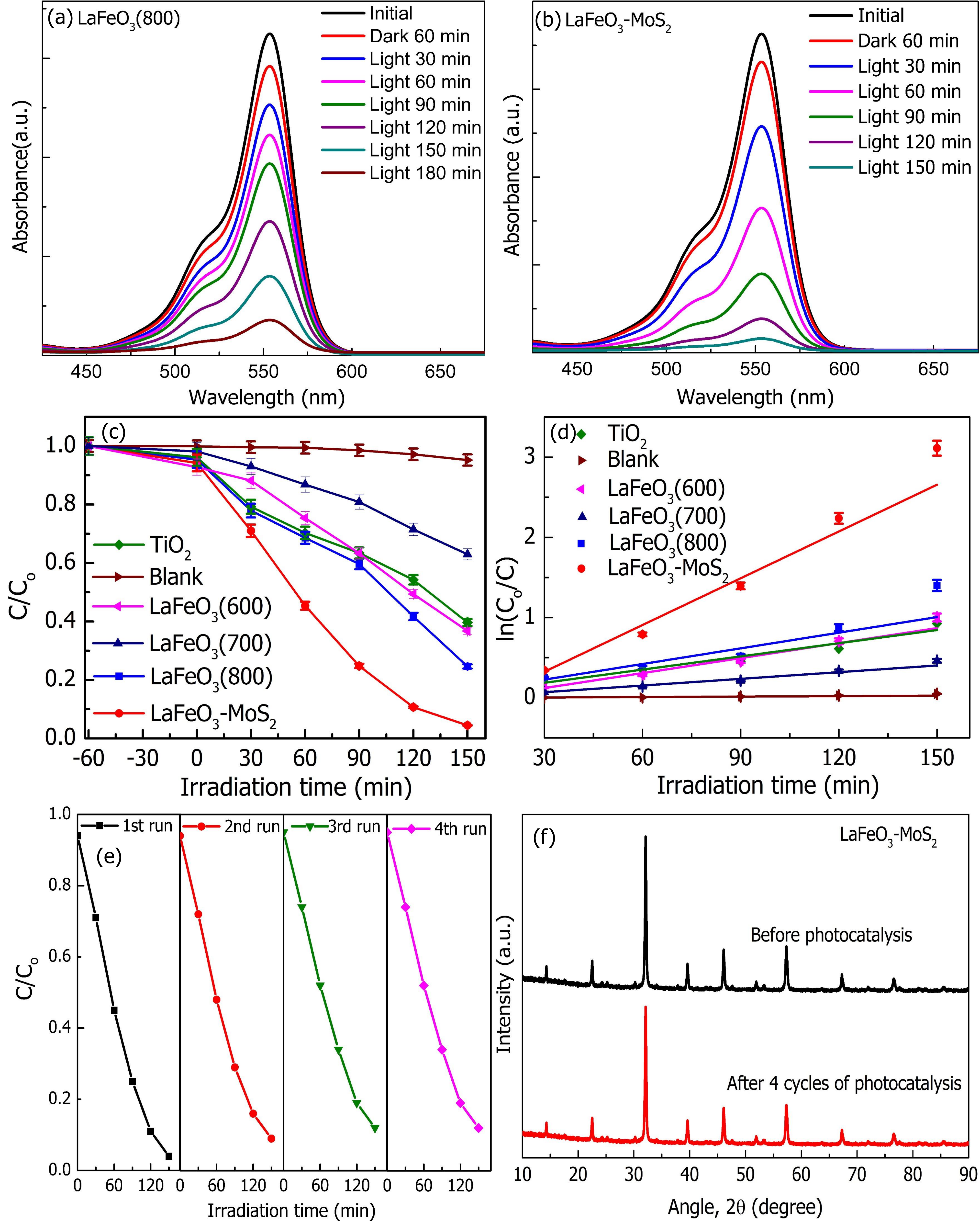}
 \caption[centering]{Spectral changes during the decolorization of RhB over (a) LaFeO\textsubscript{3} nanoparticles calcined at 800 $^{\circ}$C and (b) LaFeO\textsubscript{3}-MoS\textsubscript{2} nanocomposite under solar irradiation. (c) RhB concentration change as a function of irradiation time. (d) First-order kinetic plots for the photodegradation of RhB. (e) Recyclability of LaFeO\textsubscript{3}-MoS\textsubscript{2} nanocomposite for 4 successive runs. (f) XRD patterns of LaFeO\textsubscript{3}-MoS\textsubscript{2} nanocomposite before and after four cycles of photocatalytic experiments}
\end{figure*}

The photocatalytic efficiency of the synthesized catalysts was evaluated towards the degradation of a well-known cationic dye, RhB under solar illumination \cite{Basith2}. Initially, we had carried out the experiment at different pH and found out that the degradation percentage was increased at acidic reaction solution. Hence, pH= 3 was taken as the optimum pH to study the photochemical reactions. Further, a blank test was performed to examine the nature of RhB dye under our specified experimental conditions. Only $\sim$4\% of RhB was found to be decomposed after 150 minutes of irradiation, suggesting its negligible self-photolysis potential.  Additionally, prior to irradiation, dark adsorption tests were carried out for all the photocatalysts which demonstrated a nominal amount of RhB ($\sim$4\%) degradation after 1 hour at dark due to chemi-adsorption of the dye molecules on the surface of the photocatalysts. Therefore, considering the outcomes of these two sets of experiments, the decomposition of RhB under solar irradiation can be attributed almost entirely to photocatalysis.

\tab Fig. 5(a) and (b) present the change in the visible absorbance spectra of RhB photodegraded over LaFeO\textsubscript{3}(800) nanoparticles and LaFeO\textsubscript{3}-MoS\textsubscript{2} nanocomposite, respectively. Here, the gradual decrement in the intensity of the characteristic absorption peak at 553 nm indicates the successful decolorization of RhB via photochemical reactions. Clearly, among the two samples, the peak decreased more rapidly and took less time to ultimately disappear in case of LaFeO\textsubscript{3}-MoS\textsubscript{2} which implies its higher photocatalytic efficiency.

\subsubsection{Kinetics of RhB degradation.}

Fig. 5(c) demonstrates the degradation profiles of RhB as a function of time under both dark and light conditions in the presence of as-prepared LaFeO\textsubscript{3}(600), LaFeO\textsubscript{3}(700), LaFeO\textsubscript{3}(800) and LaFeO\textsubscript{3}-MoS\textsubscript{2} photocatalysts as well as commercially available TiO\textsubscript{2} nanoparticles. The result of blank test has also been included in the figure. The percentage of degradation has been calculated using eqn. (2)-
\begin{equation}
    Photocatalytic\;\;degradation (\%)= \frac{C_{0}-C}{C_{0}}\times 100
\end{equation}

Here, C\textsubscript{0} and C denote the initial and residual concentrations measured at 30 minutes of interval, respectively. It was found that among the three samples of LaFeO\textsubscript{3} nanoparticles, LaFeO\textsubscript{3}(800) nanoparticles have manifested the best photocatalytic performance which might be associated with its highest degree of crystallinity \cite{Othman}. However, it should be noted that despite possessing better crystallinity, LaFeO\textsubscript{3}(700) nanoparticles showed lower photocatalytic efficiency as compared to LaFeO\textsubscript{3}(600) nanoparticles which suggests that along with crystallinity other factors i.e. crystallite size, surface area may also influence the photocatalytic performance.\\

As can be seen in Fig. 5(c), LaFeO\textsubscript{3}(800) nanoparticles degraded $\sim$75\% of RhB dye after 150 minutes of illumination which was reasonably higher than that of ($\sim$60\%) TiO\textsubscript{2} nanoparticles. Interestingly, the degradation percentage of RhB increased significantly to $\sim$96\% when LaFeO\textsubscript{3}-MoS\textsubscript{2} nanocomposite was employed as the photocatalyst under the same experimental conditions. To further assess the kinetics of RhB degradation, the experimental data were fitted to zero-order (eqn. (3)), first-order (eqn. (4)) and second-order models (eqn. (5)). 

\begin{equation}
    C=-k_{0}t+C_{0}
\end{equation}
\begin{equation}
    ln(C\textsubscript{0}/C)=k_{1}t
\end{equation}
\begin{equation}
    1/C=1/C_{0}+ k_{2}t
\end{equation}

Here, t denotes illumination time and k is the overall photodegradation rate constant \cite{Mishra}. The obtained coefficient of determination ($R^{2}$) and average relative errors revealed that RhB degradation over the photocatalysts followed the first-order kinetic model. In Fig. 5(d), the slope of the linear fit plots provide the value of k\textsubscript{1} for respective photocatalysts. The rate constant in the first order kinetics of LaFeO\textsubscript{3}-MoS\textsubscript{2} nanocomposite was found to be 0.02329 min\textsuperscript{-1}, which was almost 3 and 4.25 times higher than that of the LaFeO\textsubscript{3}(800) nanoparticles (k\textsubscript{1} = 0.00932 min\textsuperscript{-1}) and TiO\textsubscript{2} nanoparticles (k\textsubscript{1} = 0.00548 min\textsuperscript{-1}), respectively. Hence, LaFeO\textsubscript{3}-MoS\textsubscript{2} nanocomposite can be considered as a superior photocatalyst as compared to the other two samples.

\begin{figure*}[t!]

 \centering
 
 \includegraphics[width= 16 cm]{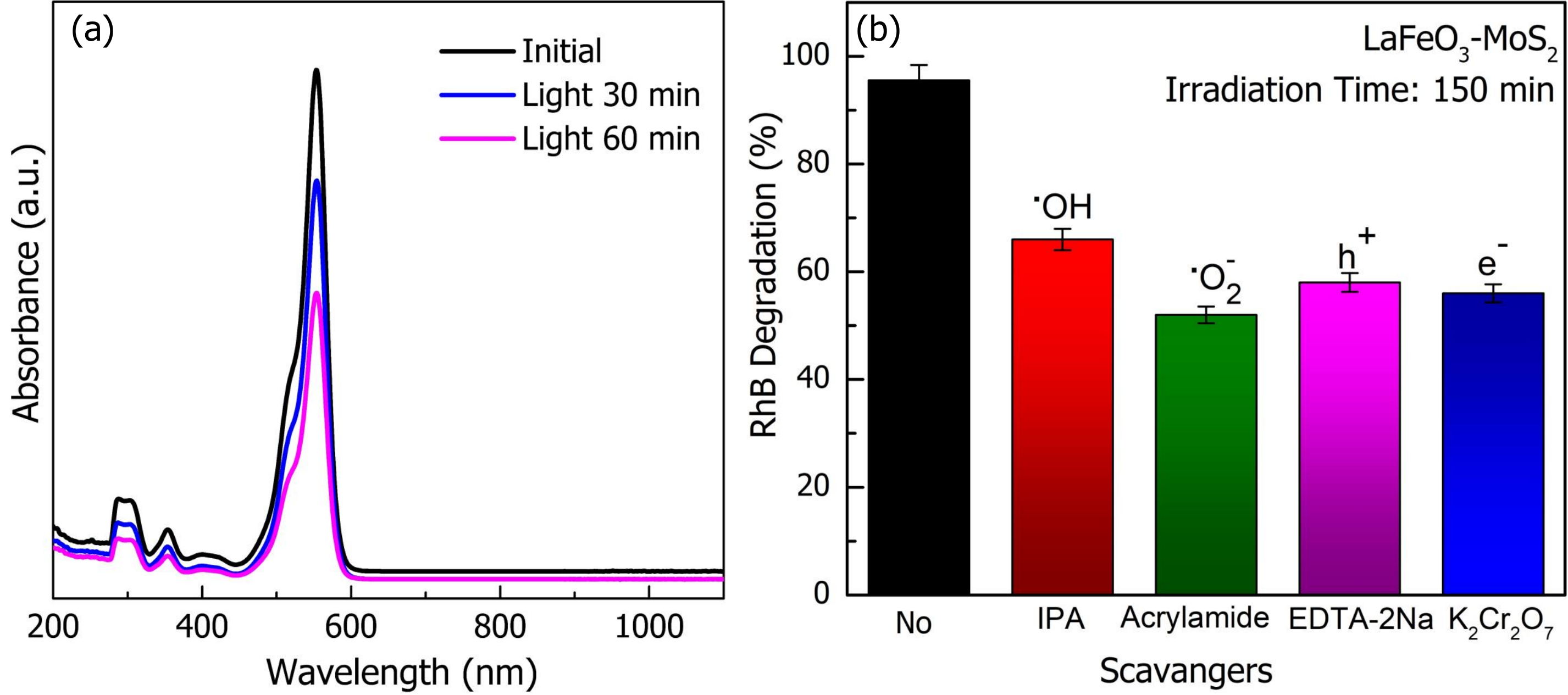}
 \caption{(a) UV-visible absorption spectra of RhB before and after being photodegraded by LaFeO\textsubscript{3}-MoS\textsubscript{2} nanocomposite under solar irradiation. (b) Effect of different scavengers on the degradation of RhB over LaFeO\textsubscript{3}-MoS\textsubscript{2} under solar irradiation.}
\end{figure*}

\subsubsection{Photostability and non-toxicity study}
To evaluate the potential for industrial-scale application, we assessed the photostability of LaFeO\textsubscript{3}-MoS\textsubscript{2} nanocomposite by performing a recyclability test \cite{Basith1, Mishra}. As shown in Fig. 5(e), the photocatalytic efficiency of the nanocomposite only slightly reduced after 4 consecutive photodegradation cycles which can be imputed to the inevitable loss of the materials during the recycling process. Furthermore, a comparison of the XRD patterns of the LaFeO\textsubscript{3}-MoS\textsubscript{2} nanocomposite before and after the 4 photocatalytic cycles is shown in Fig. 5(f). No apparent change was observed in the XRD patterns before and after the experiments, suggesting that the nanocomposite did not undergo any structural deformation during the photocatalysis. Finally, we inspected whether any toxic degradation products were evolved in the reaction mixture. For this, the UV-visible absorbance spectra (i.e. ranging from 200 nm to 1100 nm) of RhB  were examined before and after being photodegraded by LaFeO\textsubscript{3}-MoS\textsubscript{2}. As evident from Fig. 6(a), no additional peaks have emerged after illumination which implies that the photocatalyst did not introduce any carbonaceous molecules or benzene radicals in the solution. Therefore, our as-synthesized LaFeO\textsubscript{3}-MoS\textsubscript{2} nanocomposite can be considered as a promising candidate for the remediation of industrial dye effluents owing to its high photocatalytic efficiency, excellent photostability, reusability and non-toxicity. 

\subsubsection{Effect of scavengers}

To determine the major reactive species responsible for RhB degradation over LaFeO\textsubscript{3}-MoS\textsubscript{2} nanocomposite, we have performed trapping experiments using IPA, acrylamide, EDTA-2Na and K\textsubscript{2}Cr\textsubscript{2}O\textsubscript{7} as scavengers of hydroxyl ($\cdot OH $) radicals, superoxide ($ \cdot O_{2}^{-} $) radicals, holes ($h^{+}$) and electrons ($e^{-}$), respectively \cite{Acharya2017, Tama}. As evident from Fig. 6(b), the addition of acrylamide significantly decreased the degradation efficiency, i.e. from 96\% to 50\%, suggesting that $ \cdot O_{2}^{-} $ radicals were the dominant active species. Moreover, it was found that the $\cdot OH $ radicals, $h^{+}$ and $e^{-}$ also had considerable effect on the phototcatalytic process since IPA, EDTA-2Na and K\textsubscript{2}Cr\textsubscript{2}O\textsubscript{7} had reasonably reduced the degradation of RhB.

\subsection{Photodegradation of colorless Ciprofloxacin in aqueous medium}

\begin{figure*}[t!]

 \centering
 
 \includegraphics[width= 16 cm]{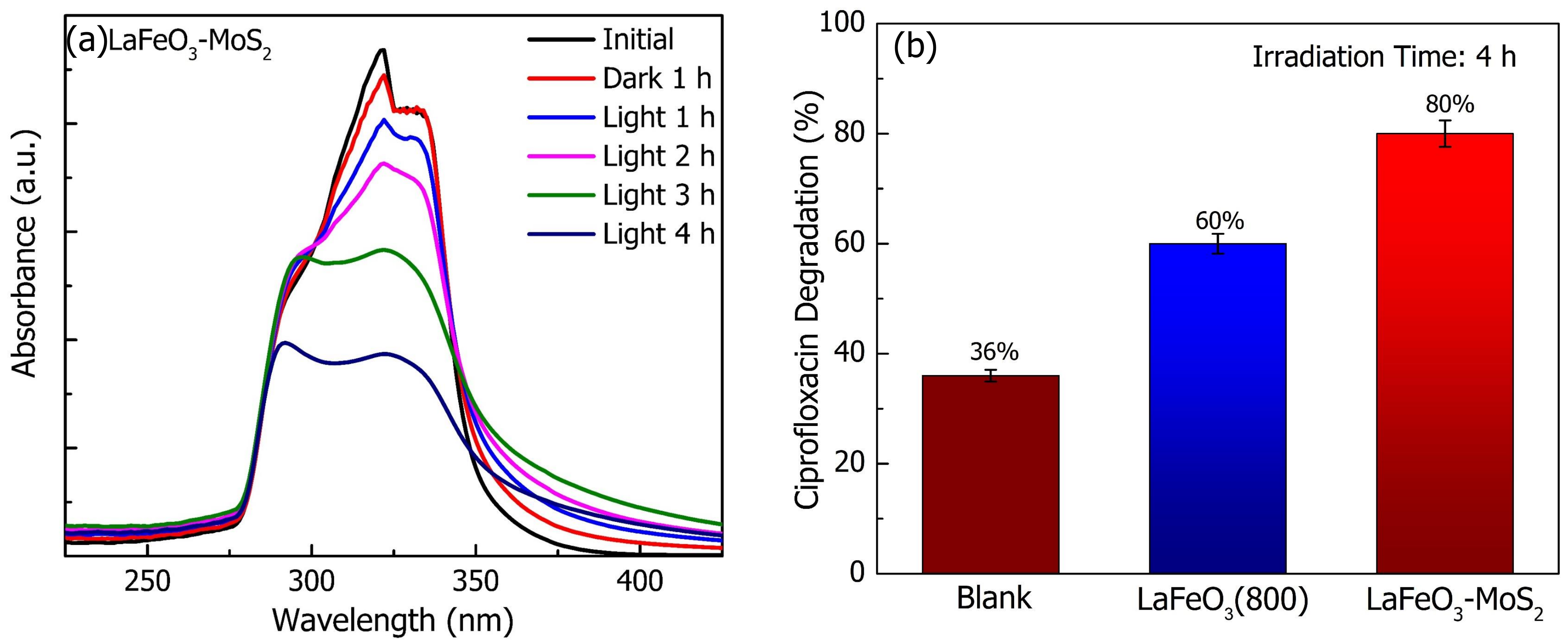}
 \caption{(a) Spectral changes during the photodegradation of ciprofloxacin over LaFeO\textsubscript{3}-MoS\textsubscript{2} nanocomposite. (b) Degradation percentage of ciprofloxacin after 4 hours of irradiation.}
\end{figure*}

\tab Fig. 7(a) demonstrates the spectral presentation of ciprofloxacin degradation in the presence of LaFeO\textsubscript{3}-MoS\textsubscript{2} nanocomposite under solar illumination. Clearly, the characteristic absorption peaks at 280 and 320 nm gradually decreased, indicating successful decomposition of ciprofloxacin \cite{Tama}. As shown in Fig. 7(b), after 4 hours of irradiation, 80\% of ciprofloxacin was photodegraded by LaFeO\textsubscript{3}-MoS\textsubscript{2} which is reasonably higher as compared to LaFeO\textsubscript{3}(800) nanoparticles (60\%). The kinetics for the decomposition of ciprofloxacin was observed to be first order (see ESI Fig. S5). The blank ciprofloxacin had a very small degradation rate (k\textsubscript{1} =0.00197 min\textsuperscript{-1}), implying its low self-degradation potential. The rate constant for LaFeO\textsubscript{3}-MoS\textsubscript{2} (k\textsubscript{1}=0.00808 min\textsuperscript{-1}) is found to be almost 2 fold higher than LaFeO\textsubscript{3}(800) (k\textsubscript{1}=0.00445 min\textsuperscript{-1}) which further justifies the superiority of the synthesized nanocomposite photocatalyst.\\
\tab Therefore, it is evident that the photocatalytic ciprofloxacin degradation efficiency of LaFeO\textsubscript{3} was significantly increased due to incorporation of few-layer MoS\textsubscript{2} nanosheets. It is worth noting that a previous investigation \cite{Sora} has reported 38\% and 90\% photocatalytic degradation of ciprofloxacin after 5 hours of solar irradiation in the presence of LaFeO\textsubscript{3} and LaFeO\textsubscript{3}+H\textsubscript{2}O\textsubscript{2}, respectively. Notably, ingestion of H\textsubscript{2}O\textsubscript{2} exerts toxic effects on human cells and results in morbidity via corrosive damage, oxygen gas formation and lipid peroxidation \cite{Watt}. Without using such a toxic reagent, we could achieve as high as 80\% degradation of ciprofloxacin employing our as-prepared LaFeO\textsubscript{3}-MoS\textsubscript{2} photocatalyst.

\subsection{Photocatalytic hydrogen generation}

\begin{figure}[t!]
 \centering
 \includegraphics[width= 8.5 cm]{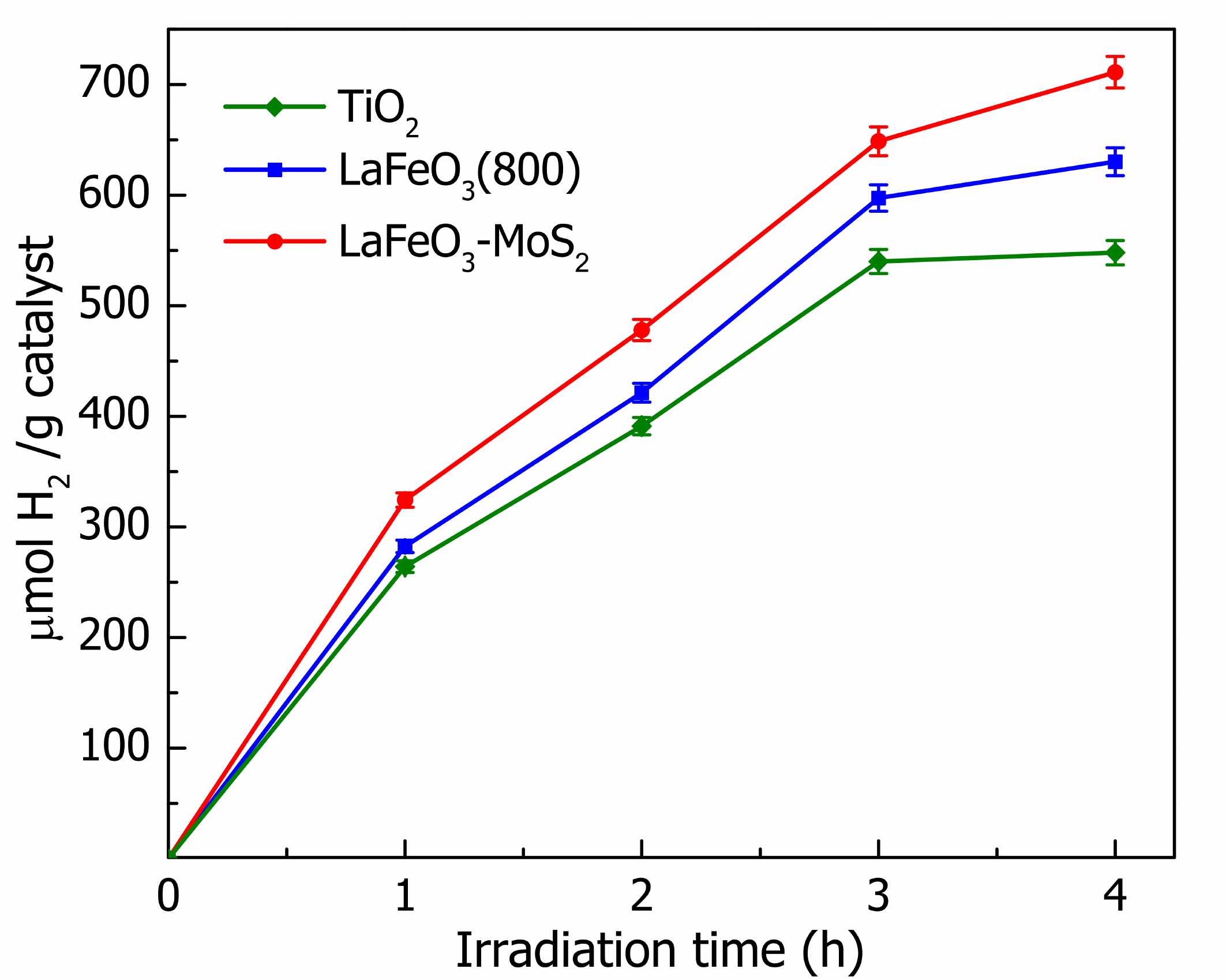}
 \caption{Amount of evolved H\textsubscript{2} via water
splitting as a function of irradiation time. 
}
\end{figure}

Photocatalytic H\textsubscript{2} generation potential of as-synthesized nanomaterials were evaluated through water splitting under solar illumination \cite{Basith2}. Fig. 8 demonstrates the amount of produced H\textsubscript{2} gas by LaFeO\textsubscript{3}(800), LaFeO\textsubscript{3}-MoS\textsubscript{2} and commercial TiO\textsubscript{2} photocatalysts as a function of time for 4 hours of irradiation. From the figure, the H\textsubscript{2} evolution rate (HER) for LaFeO\textsubscript{3}-MoS\textsubscript{2} nanocomposite has been calculated to be  $\sim$179 $\mu$mol g\textsuperscript{-1} h\textsuperscript{-1}, which is higher than that of both LaFeO\textsubscript{3}(800)  ($\sim$157.5  $\mu$mol g\textsuperscript{-1} h\textsuperscript{-1}) and TiO\textsubscript{2} nanoparticles ($\sim$137  $\mu$mol g\textsuperscript{-1} h\textsuperscript{-1}). The enhanced H\textsubscript{2} evolution potential can be attributed to the abundant catalytically active edge cites provided by few-layer MoS\textsubscript{2} nanosheets as well as increased interfacial charge transfer efficiency and high oxidation and reduction potential (will be discussed later) of the nanocomposite \cite{Zhang, Jaramillo}. \\
\tab For further insight, we have carried out a comparative analysis between the outcome of our experiments and some related previous investigations which is presented in Table 2. It is well known that photochemical reactions depend highly on experimental conditions. Therefore, for rational comparison, we have also provided details regarding light source and sacrificial reagents used in the investigations. Noticeably, the H\textsubscript{2} evolution rate (HER) of our as-prepared nanocomposite is reasonably higher than that of different LaFeO\textsubscript{3} based nanocomposites and alkaline-earth doped LaFeO\textsubscript{3} nanoparticles \cite{Xu, Vijayaraghavan}. For instance, Xu \textit{et al.} have reported that maximum HERs of 7 and 121.2  $\mu$mol g\textsuperscript{-1} h\textsuperscript{-1} can be achieved by modifying LaFeO\textsubscript{3}/g-C\textsubscript{3}N\textsubscript{4} heterojunctions with CdS and NiS co-catalysts, respectively \cite{Xu}. However, the reported synthesis route involved more precursors and complex steps as compared to our adopted technique. Additionally, our synthesized LaFeO\textsubscript{3}-MoS\textsubscript{2} has also shown significantly higher HER than a number of similar heterostructured photocatalysts such as MoS\textsubscript{2}/TiO\textsubscript{2}, MoS\textsubscript{2}/RGO, MoS\textsubscript{2}/YVO\textsubscript{4} etc \cite{Chen, Kanda, Min}. It is noteworthy that all the referred previous studies had used sacrificial reagents to facilitate the catalytic reactions, whereas our as-prepared LaFeO\textsubscript{3}-MoS\textsubscript{2} nanocomposite has demonstrated satisfactory performance in H\textsubscript{2} evolution without using any additional reagents or external energy input. Therefore, the present investigation might open-up a new avenue for low-cost solar H\textsubscript{2} production by water splitting.    

\begin{table*}[]
\caption{Comparison between the hydrogen evolution rates of LaFeO\textsubscript{3}-MoS\textsubscript{2} and  related photocatalysts.}
\resizebox{\textwidth}{!}{
\begin{tabular}{lllll}
\hline
Photocatalyst                  & Light source                          & Sacrificial reagent/Co-catalyst                        & H$_{2}$ \;\;yield       & Reference    \\
                               &                                       &                                             & ($\mu$ mol g$^{-1}$ h$^{-1}$) &              \\ \hline
2\% CdS modified LaFeO$_{3}/g-C_{3}N_{4}$ & 300 W Xe lamp; $\lambda$\textgreater{}400\;\;nm                         & 10 vol. \% triethanolamine aqueous solution & 7              & \cite{Xu}            \\
2\% NiS modified LaFeO$_{3}/g-C_{3}N_{4}$ & 300 W Xe lamp; $\lambda$\textgreater{}400\;\;nm                         & 10 vol. \% triethanolamine aqueous solution & 121.2          & \cite{Xu}          \\
0.3 molar Mg doped LaFeO$_{3}$      & Medium pressure 150 W Hg visible lamp & 10 vol. \% methanol aqueous solution        & 178.4          & \cite{Vijayaraghavan}
\\
2\% Au/LaFeO$_{3}$                  & 500 W Xe lamp                         & Na$_{2}S/Na_{2}SO_{3}$ \;\;aqueous\;\; solution                & 155            & \cite{Huang}           \\
0.047 mass\% MoS$_{2}/TiO_{2}$      & 300 W Xe lamp; $\lambda$\textgreater{}300\;\;nm                       & 5\% formic acid aqueous solution            & 2.2            & \cite{Kanda}          \\
1 wt\% $MoS_{2}/TiO_{2}$               & 300 W Xe lamp AM 1.5G filter          & 20 vol. \% methanol aqueous solution               & 119.5          & \cite{Liu2013}          \\
$MoS_{2}/RGO$                       & 300 W Xe lamp; $\lambda$ $\geq$ 420\;\; nm                   & 15 vol. \% triethanolamine aqueous solution & 38.3           & \cite{Min}            \\
2.5\% $MoS_{2}/YVO_{4}$                      & 300 W Xe lamp                   & 20 vol. \% methanol and 0.002 M $H_{2}PtCl_{6}$ aqueous solution   & 134          & \cite{Chen}           \\
$LaFeO_{3}-MoS_{2}$                    & 500 W Hg-Xe lamp                      & Aqueous solution (no reagent)                & 179            & Present work \\
\hline
\end{tabular}}
\end{table*}

\subsection{Photocatalytic mechanism}

To elucidate the rationale behind the superior photocatalytic activities of LaFeO\textsubscript{3}-MoS\textsubscript{2} nanocomposite, we have critically assessed the possible mechanisms for photodegradation of RhB dye. According to the results of the Tauc plot (Fig. 4(a)) and our previous investigation \cite{Das}, the band gap ($E_{g}$) values of as-synthesized LaFeO\textsubscript{3} and few layer MoS\textsubscript{2} nanosheets are about 2.1 and 1.21 eV, respectively. Therefore, both LaFeO\textsubscript{3} and MoS\textsubscript{2} can act as photosensitizers under visible light irradiation. \\
\tab The band edge positions of the semiconductor photocatalysts can be determined using the following empirical formulas \cite{Nethercot}.
\begin{equation}
    E_{CB}=\chi -E_{c}-\frac{1}{2}E_{g}
\end{equation}
\begin{equation}
    E_{VB}=E_{CB}+E_{g}
\end{equation}
where $\chi$ is the absolute electronegativity, expressed as the geometric mean of the absolute electronegativity of the constituent atoms, which is defined as the arithmetic mean of the atomic electron affinity and the first ionization energy; $E_{c}$ is the energy of free electrons on the hydrogen scale ($\sim$4.5 eV); $E_{g}$ is the band gap of the semiconductor; $E_{CB}$ and $E_{VB}$ are the conduction and valence band potentials, respectively. The CB and VB potentials of LaFeO\textsubscript{3} are calculated as -0.03 and 2.07 eV whereas these values for MoS\textsubscript{2} nanosheets are found to be -0.14 and 1.07 eV, respectively \cite{Acharya2017, Zeng, Jia}. Considering the band gap and band edge positions, two types of photogenerated charge-carrier transfer mechanism can be predicted for this photocatalytic system: (i) the traditional type-II double charge transfer mechanism and (ii) a direct Z-scheme mechanism. As illustrated in Fig. 9(a), in the case of type-II mechanism, the photoexcited electrons will flow from the CB of MoS\textsubscript{2} to the less negative CB of LaFeO\textsubscript{3} whereas, the flow of holes will be from the VB of LaFeO\textsubscript{3} to the less positive VB of MoS\textsubscript{2}. However, the accumulated electrons in the CB of LaFeO\textsubscript{3} are incompetent to reduce surface-absorbed O\textsubscript{2} to yield $\cdot O_{2}^{-}$ radicals since the CB potential of LaFeO\textsubscript{3} (-0.03 eV) is less negative than the redox potential of O\textsubscript{2}/$\cdot O_{2}^{-}$ (-0.046 eV vs. NHE) \cite{Wu2018}. Moreover, the VB potential of MoS\textsubscript{2} is less positive than the potential at which $H_{2}O$ molecules (or surface $OH^{-}$) are oxidized to $\cdot OH$ (1.99 eV vs. NHE) \cite{Wu2018} which suggests that the holes in the VB of MoS\textsubscript{2} cannot produce $\cdot OH$ radicals in the reaction mixture. However, the trapping experiments have already confirmed that $\cdot O_{2}^{-}$ played the vital role in the photodegradation of RhB and $\cdot OH$ was one of the subsidiary reactive species, which certainly would not be the case if type-II mechanism was followed. Therefore, it can be inferred that the photocatalytic reaction over LaFeO\textsubscript{3}-MoS\textsubscript{2} nanocomposite did not occur via the double charge-transfer mechanism path.\\ 
\begin{figure*}[t!]
\centering
\includegraphics[width= 16 cm]{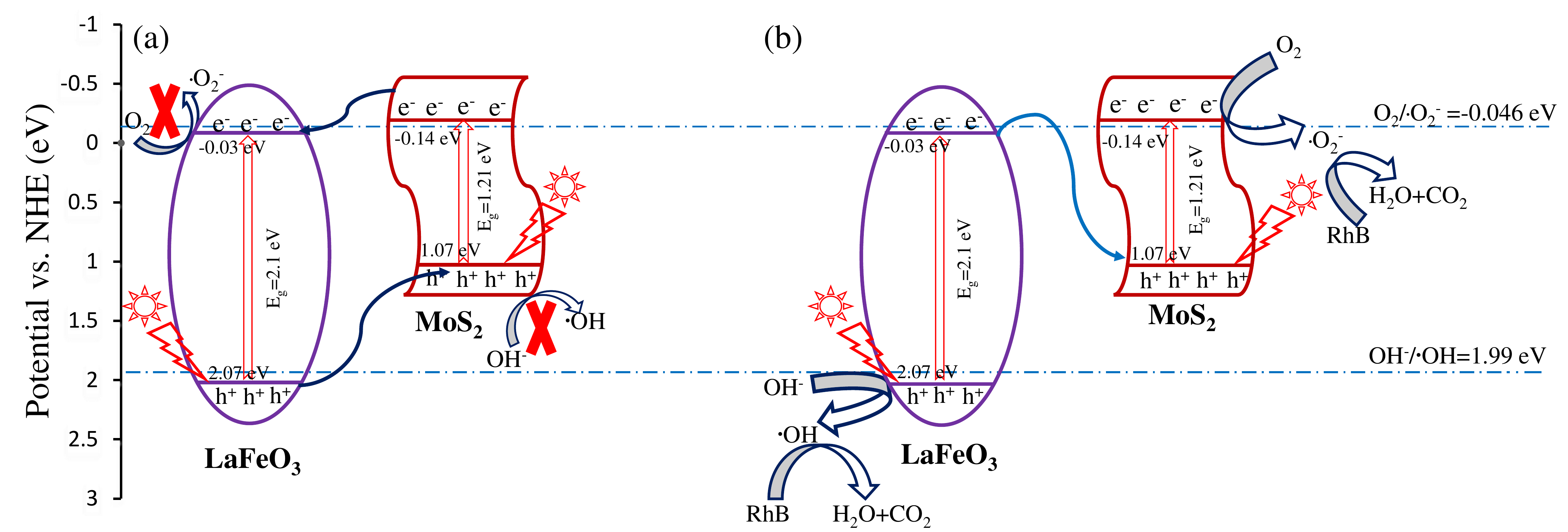}
\caption{Schematically (a) type-II double charge transfer mechanism and (b) direct Z-scheme electron transfer mechanism in the photodegradation of RhB over LaFeO\textsubscript{3}-MoS\textsubscript{2} nanocomposite under solar irradiation}
\end{figure*}
\tab The photocatalytic pathway for RhB degradation can be well explained by a mediator-free direct Z-scheme mechanism \cite{Acharya2017, Mishra}. As shown in Fig. 9(b), the photogenerated electrons from the CB of LaFeO\textsubscript{3} migrate to the more positive VB of MoS\textsubscript{2} via the heterojunction and recombine with the holes \cite{Acharya2017}. Then, the left behind electrons in the CB of MoS\textsubscript{2} successfully reduce O\textsubscript{2} to $\cdot O_{2}^{-}$ radicals since the CB potential of MoS\textsubscript{2} is more negative than the redox potential of this half-reaction. Similarly, the holes assembled in the VB of LaFeO\textsubscript{3} react with $H_{2}O$ molecules and yield $\cdot OH$ radicals owing to their sufficiently high positive potential to perform this oxidation. Further, the produced $\cdot O_{2}^{-}$ and $\cdot OH$  radicals along with the photogenerated electrons and holes react with RhB molecules and cause degradation and thereby, yield non-toxic degradation products. Rationally, all these statements are also  applicable for ciprofloxacin degradation over LaFeO\textsubscript{3}-MoS\textsubscript{2} nanocomposite. The proposed Z-scheme mechanism for photodegradation of RhB dye over LaFeO\textsubscript{3}-MoS\textsubscript{2} nanocomposite can be summarized as follows:
\begin{equation}
    LaFeO_{3}+h\nu \rightarrow LaFeO_{3}(e^{-}+h^{+})
\end{equation}
\begin{equation}
     MoS_{2}+h\nu \rightarrow MoS_{2}(e^{-}+h^{+})
\end{equation}
\begin{equation}
    LaFeO_{3}-MoS_{2}(e^{-}+h^{+}) \rightarrow LaFeO_{3}(h^{+})+MoS_{2}(e^{-})
\end{equation}
\begin{equation}
    LaFeO_{3}(h^{+}) + H_{2}O \rightarrow H^{+}+\cdot OH
\end{equation}
\begin{equation}
    MoS_{2}(e^{-})+O_{2} \rightarrow \cdot O_{2}^{-}
\end{equation}
\begin{equation}
    RhB+(\cdot O_{2}^{-},\cdot OH,e^{-},h^{+}) \rightarrow Degradation\;\;products
\end{equation}

The results of photocatalytic $H_{2}$ evolution tests further substantiate that LaFeO\textsubscript{3}-MoS\textsubscript{2} photocatalyst is more likely to follow direct Z-scheme mechanism, rather than type-II double charge transfer mechanism. Both mechanisms for solar $H_{2}$ generation are schematically presented in ESI Fig. S6. According to thermodynamic principles, for successful water splitting, the photogenerated electrons must have more negative potential than the reduction potential of $H^{+}$/$H_{2}$ (0 eV vs. NHE) and simultaneously, the holes must be more positive than the oxidation potential of $H_{2}O$/$O_{2}$ (1.23 eV vs. NHE) \cite{Yendrapati}. Therefore, if type-II mechanism would have existed, LaFeO\textsubscript{3}-MoS\textsubscript{2} photocatalyst could not generate $H_{2}$ via water splitting (see ESI Fig. S6(a)) since the VB potential of MoS\textsubscript{2} is less positive than the oxidation potential of $H_{2}O$/$O_{2}$. Contrarily, in the case of direct Z-scheme mechanism, the holes in the VB of LaFeO\textsubscript{3} can produce $O_{2}$ and $H^{+}$ by oxidizing the surface-adsorbed $H_{2}O$ molecules owing to their adequately high redox potential. Likewise, the electrons accumulated in the CB of MoS\textsubscript{2} will also be able to successfully reduce $H^{+}$ and evolve $H_{2}$ gas. Hence, it can be inferred that LaFeO\textsubscript{3}-MoS\textsubscript{2} photocatalyst has produced solar $H_{2}$ via direct Z-scheme mechanism. Moreover, the unsaturated active S atoms with mono- or two-coordination on exposed edges of MoS\textsubscript{2} nanosheets might have strong bonds with the $H^{+}$ ions present in the solution and thus, can easily capture them \cite{Chen, Tama}. As a consequence, the reduction of $H^{+}$ to $H_{2}$ would have been accelerated, resulting in high HER which is consistent with our experimental findings. \\
\tab Based on the outcomes and above analysis, the superior photocatalytic performance of as-synthesized direct Z-scheme LaFeO\textsubscript{3}-MoS\textsubscript{2} photocatalyst can be attributed to its enhanced charge separation and transportation efficiency at heterojunction as well as high oxidation and reduction potentials of photoinduced holes and electrons. In addition to these, the excellent crystallinity, favorable surface morphology and improved ferroelectric behavior of the nanocomposite after the incorporation of MoS\textsubscript{2} nanosheets might have played significant role to enhance the photocatalytic efficiency \cite{Basith1}.

\section{Conclusions}

We observed that the as-prepared LaFeO\textsubscript{3}-MoS\textsubscript{2} photocatalyst has considerably higher degradation efficiency of RhB as compared to pristine LaFeO\textsubscript{3} and commercially available titania nanoparticles. Moreover, the superior performance of the nanocomposite towards the photodecomposition of a colorless micropollutant has ensured that the degradation is solely attributed to photocatalytic process rather than dye-sensitization effect. Such enhanced photocatalytic performance of this nanocomposite can be mainly ascribed to the formation of a direct Z-scheme heterojunction between its constituents which facilitated charge separation via interfacial interaction and provided charges with strong redox ability to drive photocatalysis. Further, the plethora of exposed active edge sites on the surface of lamellar MoS\textsubscript{2} might effectively promote the H\textsubscript{2} evolution reaction, resulting in significant amount of H\textsubscript{2} generation via water splitting. It is worth mentioning that the LaFeO\textsubscript{3}-MoS\textsubscript{2} nanocomposite with high photocatalytic efficiency, non-toxicity and excellent recyclability, can be considered as a promising photocatalyst for numerous practical applications such as removal of harmful textile and pharmaceutical pollutants from wastewater, solar water disinfection, CO\textsubscript{2} reduction, low-cost photocatalytic H\textsubscript{2} evolution etc. The present investigation might also open up new avenues for the design and synthesis of novel 2D transition-metal dichalcogenides based Z-scheme photocatalysts for environmental remediation, solar energy conversion and related applications.

\section*{Acknowledgements}
We acknowledge the Committee for Advanced Studies and Research (CASR), Bangladesh University of Engineering and Technology for financial assistance. This work was financially supported partly by the Ministry of Education, Government of Bangladesh (Grant No. 37.20.0000.004.033.020.2016./PS20191017).
\section*{Appendix A. Supplementary data} 
 Supplementary data to this article can be found online. See DOI:\href{https://doi.org/10.1016/j.mseb.2021.115295}{10.1016/j.mseb.2021.115295}

\section*{Data availability}
The raw and processed data required to reproduce these findings cannot be shared at this time due to technical or time limitations.








\end{document}